\documentclass[useAMS,usenatbib]{mn2e}
\usepackage{latexsym,graphicx,natbib,amssymb}

\title[The role of collisions for massive star formation]
 {The role of stellar collisions for the formation of massive stars}
      
\author[Baumgardt \& Klessen]
{{\Large H. Baumgardt$^{1}$ and R. S. Klessen$^{2,3}$}\\
 $^1$ School of Mathematics and Physics, The University of Queensland, Brisbane, QLD 4072, Australia \\ 
 $^2$ Zentrum f\"ur Astronomie der Universit\"at Heidelberg, Institut f\"ur Theoretische Astrophysik, Albert-Ueberle-Str. 2, 69120     Heidelberg, Germany \\ 
 $^3$ Kavli Institute for Particle Astrophysics and Cosmology, 
Stanford University, Menlo Park, CA 94025, U.S.A.
}

\date{Accepted ????. Received ?????; in original form ?????}

\pagerange{\pageref{firstpage}--\pageref{lastpage}} \pubyear{2008}

\begin{document}   

\maketitle

\label{firstpage}
 
\begin{abstract}
We use direct $N$-body simulations of gas embedded star clusters to study the importance 
of stellar collisions for the formation and mass
accretion history of high-mass stars. Our clusters start in virial equilibrium as a mix 
of gas and proto-stars. Proto-stars then accrete matter using different mass accretion rates
and the amount of gas is reduced in the same way as the mass of stars increases. During the 
simulations we check for stellar collisions and we investigate the role of these collisions 
for the build-up of high-mass stars and the formation of runaway stars. 

We find that a significant number of collisions only occur in clusters with initial half-mass radii 
$r_h \le 0.1$ pc. After emerging from their parental gas clouds, such clusters end up too 
compact compared to observed young, massive open clusters. In addition, collisions lead mainly to the formation 
of a single runaway star instead of the formation of many high mass stars with a broad mass spectrum. We 
therefore conclude that massive stars form mainly by gas accretion, with stellar collisions only playing 
a minor role if any at all. 
Collisions of stars in the pre-main sequence phase might however contribute to the formation of the most massive 
stars in the densest star clusters and possibly to the formation of intermediate-mass
black holes with masses up to a few 100 M$_\odot$. 
\end{abstract}


\section{Introduction}
\label{sec:intro}

An important unanswered question in modern astrophysics concerns the mechanism(s) by which stars and star clusters
form. This is especially true in the case of massive stars, which are challenging objects to observe since they are 
rare, evolve on short timescales, and are often highly extincted. Various theories have been proposed concerning the 
formation process of massive stars, like monolithic collapse of molecular cloud cores (McKee \& Tan 2003), 
competitive gas accretion in small clusters (Larson 1978, Bonnell et al. 1997) driven by turbulent fragmentation
({Klessen \& Burkert 2000,} Klessen, Heitsch \& Mac Low 2000, Klessen 2001) and stellar collisions and mergers in dense clusters 
(Bonnell et al. 1998, Clarke \& Bonnell 2008). For further discussions, see Mac~Low \& Klessen (2004), McKee \& Ostriker (2007), and Zinnecker \& Yorke (2007).

Calculations of the collapse of protostellar clouds show that the collapse proceeds non-homologously and in such a way
that the central regions rapidly form a proto-stellar core while the outer regions still continue to fall inward 
(Bodenheimer \& Sweigart 1968, Larson, 1969, Shu 1977). The formation of low and intermediate-mass stars ($M \le 10$ M$_\odot$)
can readily be explained this way (Palla \& Stahler 1993), however this models fails for high-mass stars since
for stars more massive than about 20 M$_\odot$ the Kelvin-Helmholtz timescale is larger than their formation time,
meaning that these stars start core hydrogen burning while they are still accreting. Once the resulting radiation pressure 
from the stellar core exceeds the gravitational attraction, further accretion of gas is halted, limiting stellar masses to
20 M$_\odot$ at least for spherically symmetric accretion (Kahn 1974, Wolfire \& Cassinelli 1987). The 
effects of non-ionizing as well ionizing radiation, however,
might be overcome if accretion happens through
an accretion disc which shields the gas from radiation (Yorke \& Sonnhalter 2002, Krumholz et al. 2009, Peters et al. 2010a,b,c).   

Based on the fact that massive stars are mainly found in the central regions of rich star clusters 
(Hillenbrand 1997, Hillenbrand \& Hartmann 1998), Bonnell et al. (1998) presented a model for high-mass star 
formation in which high-mass stars form through gas accretion and repeated mergers in the center of a star cluster. 
In their calculations, the clusters
contracted while stars were accreting until the central density became high enough for stellar collisions to occur. 
Once a massive star has formed, the remaining cluster gas is removed and the cluster expands again.
Clarke \& Bonnell (2008) found furthermore that relaxation effects between proto-stars prevent low-mass clusters from
reaching high-enough densities needed for collisions, in good agreement with the fact that high-mass stars are only
found in massive clusters \citep{wk06}.

Direct simulations of the role of stellar collisions for star formation are still lacking. In the present paper we present 
$N$-body simulations of the early evolution of proto-clusters.
Our clusters start as a mix of gas and proto-stars and we follow the cluster evolution through the accretion and
gas expulsion phases. Our paper is organized as follows: In \S\ \ref{sec:mod} we describe the details of our 
simulations. In \S\ \ref{sec:results} we present our results and in
\S\ \ref{sec:concl} we draw our conclusions.

\section{Models}
\label{sec:mod}

All simulations were made with the graphics-card 
enabled version of the collisional $N$-body code NBODY6
\citep{a99} on NVIDIA graphic processing units (GPU). NBODY6 uses a Hermite scheme to integrate the motion of stars,
and uses an Ahmad-Cohen type neighbor scheme \citep{ac73} for the calculation of forces from the nearest neighbors.
It also uses KS \citep{ks65} and Chain regularisation \citep{ma90} to deal with close encounters between stars. 

In our simulations, the star clusters were assumed to be initially a mix of gas and accreting proto-stars.
The gas was not simulated directly, instead its influence on the stars was modeled as a modification to the 
equation of motion of stars.  We assumed that the star formation efficiency does not depend on the position inside 
the cluster, so gas and stars followed the same density distribution initially, which was given by a Plummer model.
The influence of the external gas was calculated at each regular time step of NBODY6. Since NBODY6 uses 
a Hermite scheme to integrate the motion of stars, the correction terms due to
the external gas on the acceleration and its first derivative were evaluated at each regular step.
More details on the way the external gas was implemented can be found in \citet{bk07}.

The overall star formation efficiency was set to be 30\%, i.e. 70\% of the initial gas was not converted into stars,
which agrees with typical star formation efficiencies found for embedded clusters in the Milky Way \citep{ll03}.
We assumed that the removal of gas not converted into stars sets in after $t_d=0.2$ Myr or, if the formation timescale
is longer, after the most massive stars have formed. After this time, the fraction of gas was assumed to decrease 
exponentially on a timescale of $\tau_M=0.5$ Myr, i.e.
\begin{equation}
 M_{gas}(t) = M_{gas tot} \; e^{-(t-t_D)/\tau_M} \; \; .
\end{equation}
The timescale for $\tau_M$ was chosen such that our clusters are gas free after a few Myr, in agreement with observations
which show that most embedded star clusters have ages of less than a few Myr (Lada \& Lada 2003). All simulations were run 
for 10 Myr to allow a comparison of our simulations with observations of gas embedded star clusters and young open clusters.

All proto-stars started with a mass of $0.1$ M$_\odot$ initially. We assumed that all stars accrete with the 
same accretion rate. Most simulations were done with an accretion rate of $\dot{M}=10^{-4}$ M$_\odot$/yr, but we  
also made simulations with accretion rates of $\dot{M}=10^{-5}$ M$_\odot$/yr and $\dot{M}=10^{-3}$ M$_\odot$/yr
to test the influence of the assumed accretion rate on our results. It is possible that  
the accretion rate onto massive stars is higher than onto low-mass stars, as e.g. indicated by infall motions 
and other accretion tracers (e.g. Fuller et al. 2005, Keto \& Wood 2006). Our simulations should therefore be regarded 
as first steps to address the question of stellar collisions during star formation. Due to the constant accretion rate for all 
stars, low-mass stars reach their
final mass first, in agreement with observations of young open clusters in the Milky Way, which show that high mass 
stars are still accreting after low-mass stars have completed their accretion \citep[e.g.][]{h62,detal01,kkc06}. 
The mass of the gas was reduced in the same way as the stars gained mass such that
the total cluster mass was constant before the onset of gas removal.

Accretion onto the protostars was stopped once the mass of each star reached a pre-specified final  
mass. The mass function of these final masses was given by a Kroupa IMF \citep{k01} between lower and upper mass
limits of $m_{low}=0.1$ M$_\odot$ and $m_{up} = 15$ M$_\odot$.  Due to the low upper mass limit in our runs,
any star with mass $m>15$ M$_\odot$ cannot form directly but has to form via collisions between lower-mass stars.

The radius evolution of the stars in our simulations was split into three different
phases, the accretion phase, the cooling phase and the stellar evolution phase after stars have reached the 
main sequence. 

During the accretion phase proto-stars grow in mass until they have reached their final mass.
Following \citet{ho09}, we assume that in the accretion phase proto-stellar radii depend on the actual mass of a star and 
its accretion rate, but not on the final mass. In our simulations, the radius of a star in the accretion phase is given by:
\begin{equation}
\frac{R_*}{R_\odot} = 421 \cdot \left( \frac{\dot{M}}{{\rm M}_\odot/{\rm yr}} \right)^{0.375} \cdot \left( \frac{M}{{\rm M}_\odot} \right)^{0.33} 
\label{radacc}
\end{equation}

\begin{figure}
\begin{center}
\includegraphics[width=8.5cm]{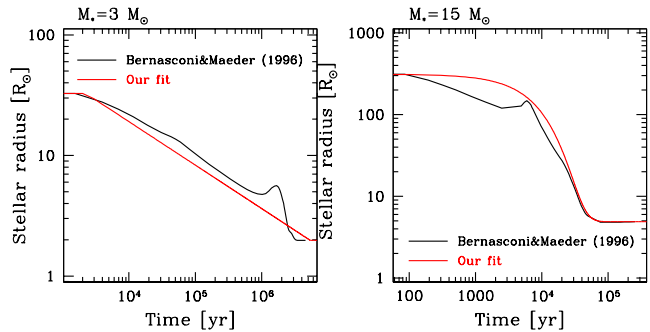}
\end{center}
\caption{Pre-main sequence evolution of stellar radius as a function of time for two different stellar masses according to the models by 
\citet{bm96} (solid line) and our fitting formulae (red dashed lines). It can be seen that our fitting formulae are within 20\% of 
the Bernasconi \& Maeder models for most of the time.} 
\label{fig:radev}
\end{figure}

In the contraction phase stars contract towards
the zero-age main sequence after having reached their final mass.
During the contraction phase, we assume that the radius of a star depends on its mass and the time which has passed since the star 
has reached 
its final mass. To roughly match the results of the detailed calculations by \citet{bm96}, we assumed that for stars more 
massive than 7 M$_\odot$, the stellar radius decreases exponentially towards the zero-age main sequence such that:
\begin{equation}
 R_* = R_{ZAMS} + (R_{max}-R_{ZAMS}) \cdot \exp (-(t-t_{max})/t_{cool}) \;\; . 
\end{equation}
Here $R_{ZAMS}$ are the zero-age main-sequence radii of stars which we have taken from \citet{tetal96}, assuming solar metallicity.
$R_{max}$ is the maximum radius reached during the accretion phase according to eq. \ref{radacc}, $t_{max}$ is the time
when this radius is reached, and $t_{cool}$ is the cooling timescale which is given by
\begin{equation}
 \log \frac{t_{cool}}{\rm yr} = 11.57 - 9.85 \cdot \log \frac{M}{{\rm M}_\odot} + 2.86 \cdot \left( \log \frac{M}{{\rm M}_\odot} \right)^2 
\end{equation}
After 5 $t_{cool}$, we assumed that stars start their main sequence evolution, which was modeled according to the fitting formulae
of \citet{hpt00}. For stars less massive than 7 M$_\odot$, the Bernasconi \& Maeder (1996) tracks differ significantly from an exponential decrease,
so we assumed a linear decrease of the stellar radius with time according to
\begin{eqnarray}
\nonumber \log \frac{R_*}{R_\odot} & = & \log R_{ZAMS} + (\log R_{max}- \log R_{ZAMS})  \\
  & &  \cdot (\log t_2 - \log (t-t_{max}))/(\log t_2 - \log t_1) \;\; ,
\end{eqnarray}
for times $t_1 < t < t_2$, where the constants $t_1$ and $t_2$ were given by 
\begin{eqnarray}
  t_1 & = & 4.677 \cdot 10^{-3} \cdot \left( \frac{M}{{\rm M}_\odot} \right)^{-2/3}  {\rm Myr}  \\
  t_2 & = & 23.4 \cdot \left( \frac{M}{{\rm M}_\odot} \right)^{-4/3}  {\rm Myr} \;\;\; . \\
\end{eqnarray}
For times $t<t_1$, we assumed $R_*=R_{max}$, while for times $t>t_2$ stars were assumed to undergo standard
stellar evolution. Fig.\ \ref{fig:radev} shows the time evolution of stellar radii for two different masses according to 
\citet{bm96} and our simple fitting formulae. We adjusted the accretion rate such that our models start at the same initial 
radius as the models of Bernasconi \& Maeder (1996). Our fitting formulae are within 20\% of the
Bernasconi \& Maeder models for most times, which should be sufficiently accurate for our purposes.

\begin{table*}
\centering
\caption[]{Details of the performed $N$-body runs}
\begin{tabular}{c@{\hspace{0.15cm}}r@{\hspace{0.15cm}}c@{\hspace{0.15cm}}c@{\hspace{0.15cm}}r@{.}l@{\hspace{0.15cm}}c@{\hspace{0.15cm}}c@{\hspace{0.15cm}}r|c@{\hspace{0.15cm}}r@{\hspace{0.15cm}}c@{\hspace{0.15cm}}l@{\hspace{0.15cm}}r@{.}l@{\hspace{0.15cm}}c@{\hspace{0.15cm}}c@{\hspace{0.15cm}}r}
\hline\hline
  \multicolumn{1}{r}{$N_{Run}$} & $N_{Star}$ & $M_{Acc}$ & Type & \multicolumn{2}{c}{$r_{h ini}$} & $\langle N_{Coll}\rangle$ & $M_{Max}$& $r_{hp fin}$ &
   $N_{Run}$ & $N_{Star}$ & $M_{Acc}$ & Type & \multicolumn{2}{c}{$r_{h ini}$} & $\langle N_{Coll} \rangle$ & $ M_{Max}$  & $r_{hp fin}$\\
  & & [$M_\odot$/yr] & & \multicolumn{2}{c}{[pc]} & & [$M_\odot$] & [pc] &  & & [$M_\odot$/yr] & & \multicolumn{2}{c}{[pc]} & & \multicolumn{1}{c}{[$M_\odot$]} & \multicolumn{1}{c}{[pc]} \\[+0.2cm]
 10 & 1000 & $10^{-4}$ & Std  & 0 & 033 & $\;\,$2.1  &  $\;\,$17.4 & 0.25 & 10 & 1000 & $10^{-4}$ & Segr & 0&033 & $\;\;\;$9.7   & $\;\;$19.0 & 2.36 \\
 10 & 3000 & $10^{-4}$ & Std  & 0 & 033 & $\;\,$5.4  &  $\;\,$25.4 & 0.20 & 10 & 3000 & $10^{-4}$ & Segr & 0&033 & $\;$43.5  & $\;\;$49.6 & 0.92 \\
 $\;\;$1 & 10000 & $10^{-4}$ & Std  & 0 & 033 & 41.0 & 139.9 & 0.22 & $\;\;$1 & 10000 & $10^{-4}$ & Segr & 0&033 & 220.0 & 802.2& 2.43 \\
 10 & 1000 & $10^{-4}$ & Std  & 0 & 10  & $\;\,$0.3  &  $\;\,$12.3 & 0.23 & 10 & 1000 & $10^{-4}$ & Segr & 0&10  & $\;\;\;$2.9   & $\;$27.6 & 0.31 \\
 10 & 3000 & $10^{-4}$ & Std  & 0 & 10  & $\;\,$1.4  &  $\;\,$20.5 & 0.25 & 10 & 3000 & $10^{-4}$ & Segr & 0&10  & $\;\;\;$9.9   & $\;$31.3 & 0.24 \\
 $\;\;$1 & 10000 & $10^{-4}$ & Std  & 0 & 10  & $\;\,$3.0  &  $\;\,$43.6 & 0.23 & $\;\;$1 & 10000 & $10^{-4}$ & Segr & 0&10  & $\;$54.0  & 272.1& 0.23 \\
 10 & 1000 & $10^{-4}$ & Std  & 0 & 33  & $\;\,$0.0  &  $\;\,$13.1 & 0.47 & 10 & 1000 & $10^{-4}$ & Segr & 0&33  & $\;\;\;$0.3   & $\;$14.6 & 0.28 \\
 10 & 3000 & $10^{-4}$ & Std  & 0 & 33  & $\;\,$0.1  &  $\;\,$13.5 & 0.51 & 10 & 3000 & $10^{-4}$ & Segr & 0&33  & $\;\;\;$2.7   & $\;$27.4 & 0.28 \\
 $\;\;$1 & 10000 & $10^{-4}$ & Std  & 0 & 33  & $\;\,$0.0  &  $\;\,$14.1 & 0.62 & $\;\;$1 & 10000 & $10^{-4}$ & Segr & 0&33  &$\;$ 11.0  & $\;$27.8 & 0.20 \\
 10 & 1000 & $10^{-4}$ & Std  & 1 & 00   & $\;\,$0.0  &  $\;\,$12.9 & 4.26 & 10 & 1000 & $10^{-4}$ & Segr & 1&00   & $\;\;\;$0.0   & $\;$13.1 & 0.32 \\
 10 & 3000 & $10^{-4}$ & Std  & 1 & 00   & $\;\,$0.0  &  $\;\,$13.3 & 4.17 & 10 & 3000 & $10^{-4}$ & Segr & 1&00   & $\;\;\;$0.6   & $\;$17.8 & 0.30 \\
 $\;\;$1 & 10000 & $10^{-4}$ & Std  & 1 & 00   & $\;\,$0.0  &  $\;\,$14.6 & 2.11 & $\;\;$1 & 10000 & $10^{-4}$ & Segr & 1&00   & $\;\;\;$1.0   & $\;$25.6 & 0.22 \\
 10 & 1000 & $10^{-4}$ & Bin  & 0 & 033 & 11.7  &  $\;\,$12.0 & 0.50 & 10 & 1000 & $10^{-3}$ & Std  & 0 & 033 & $\;\,$1.8  &  $\;\,$17.1 & 0.51 \\
 10 & 3000 & $10^{-4}$ & Bin  & 0 & 033 & 42.0  &  $\;\,$25.4 & 0.31 &  10 & 3000 & $10^{-3}$ & Std  & 0 & 033 & $\;\,$6.6  &  $\;\,$32.4 & 0.24 \\
 10 & 1000 & $10^{-4}$ & Bin  & 0 & 10  & $\;\,$1.5  &  12.8 & 0.41 &  $\;\;$1 & 10000 & $10^{-3}$ & Std  & 0 & 033 & 31.0  &  113.3 & 0.19 \\
 10 & 3000 & $10^{-4}$ & Bin  & 0 & 10  & 14.4  &  $\;\,$18.7 & 0.32 &  10 & 1000 & $10^{-5}$ & Std  & 0 & 033 & $\;\,$3.3  &  $\;\,$20.7 & 0.55 \\
 10 & 1000 & $10^{-4}$ & Bin  & 0 & 33 & $\;\,$0.1  &  $\;\,$11.8 & 0.72 &  10 & 3000 & $10^{-5}$ & Std  & 0 & 033 & 11.6  &  $\;\,$22.4 & 0.32 \\
 10 & 3000 & $10^{-4}$ & Bin  & 0 & 33 & $\;\,$0.7  &  $\;\,$13.2 & 0.25 &  $\;\;$1 & 10000 & $10^{-5}$ & Std  & 0 & 033 & 68.0  &  285.6 & 0.29 \\
 10 & 1000 & $10^{-4}$ & Bin  & 1 & 0  & $\;\,$0.0  &  $\;\,$12.1 & 3.87 & \\ 
 10 & 3000 & $10^{-4}$ & Bin  & 1 & 0  & $\;\,$0.1  &  $\;\,$14.3 & 3.63 & \\
\hline \hline
\end{tabular}
\label{tabradvel}
\end{table*}

Stars were merged if their separation became smaller than the sum of their radii and the mass of the merger
product was set equal to the sum of the masses of the stars, i.e. we assumed no disruption of stars and no mass
loss during a collision. 
Merged stars were assumed to be immediately back on their pre-main sequence or main-sequence track, so we did not allow for an 
increased radius of a star due to energy increase as a result of a collision. This is justified since thermal adjustment of massive
collision products happens on timescales of $10^3 - 10^4$ yr \citep{setal07}, which is much smaller than the typical time for collisions except
in the most compact clusters. We also did not allow for tidal interactions during close flybys or collisions which do not lead to mergers
between stars. Incorporation of these effects is difficult since they depend on the internal structure of stars and would need to be 
followed by detailed simulations of the pre-MS evolution and the collisions.

\section{Results}
\label{sec:results}

Observations of open clusters in the Milky Way show that the formation of O stars, i.e. stars with masses larger than $\sim$25 M$_\odot$ 
is happening in clusters down to about 1000 stars \citep{wkb10}, so in our simulations we followed the evolution of 
star clusters containing between $10^3$ to $10^4$ stars. Simulations with less than $10^4$ stars were repeated 10 times and 
the results were averaged to reduce the 
statistical noise of low-$N$ simulations. The initial half-mass radii of the clusters were varied between
0.033 pc to 1.0 pc. Embedded star clusters usually have radii between 0.3 to 1 pc (Lada \& Lada 2003), so our 
simulations more than cover the range of half-mass radii found for embedded star clusters. For each mass and radius, we simulated
two kinds of clusters, mass segregated and unsegregated models. In the unsegregated models, the stellar number density follows a Plummer
profile and there is no correlation between the final mass of a star and its position in the cluster. The gain in mass by the stars
is balanced by the mass loss from the gas such that the enclosed mass within any radius does not change. Hence, the Lagrangian radii
of unsegregated clusters remain almost constant in time throughout the accretion phase, except for possible relaxation effects. 

In the segregated models, the stellar number density profile still follows a Plummer profile initially, however we correlate the 
final mass of each star with its total energy in the cluster such that the most massive stars have the lowest energies and will
form mainly in the core of the cluster.
In the mass segregated models, the mass profile therefore changes during the accretion phase: Due to the high number of massive stars 
in the core, the core mass increases and its size decreases as it tries to remain in virial equilibrium. The mass-segregated simulations 
mimic the behavior of an embedded star cluster where the core mass increases due to gas which cools and falls into the center. 
They are also similar to the star clusters discussed by \citet{bbz97}, which also form a very compact core of massive stars. 
Table~1 summarizes our suite of simulations. It shows the number of runs done for any particular model, the number of stars, the assumed
mass accretion rate, the type of run (Std: Unsegregated cluster, Segr: Mass segregated cluster, Bin: Simulation with binaries), the 
initial half-mass radius, the number of collisions, and the maximum stellar mass and cluster half-mass radius at T=10 Myr.
 
\subsection{Unsegregated clusters}
\label{sec:noseg}

\begin{figure}
\begin{center}
\includegraphics[width=8.5cm]{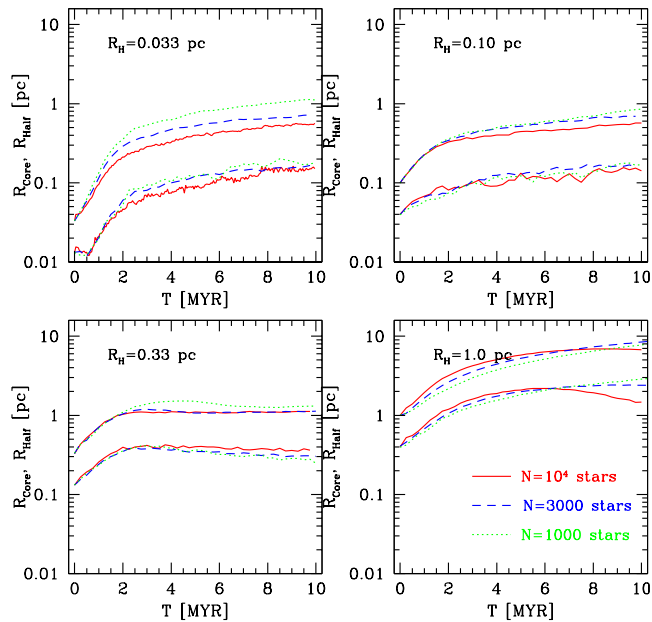}
\end{center}
\caption{Evolution of Lagrangian radii for unsegregated clusters as a function of time. Shown are half-mass radii (upper curves) and 10\% Lagrangian
radii (lower curves). Clusters expand due to gas expulsion and mass loss from massive stars. The most concentrated models reach core-collapse 
before gas expulsion sets in and expand up to a factor 10 within the first 10 Myr due to strong mass loss from the core and binary heating 
(upper left panel). Less concentrated clusters expand by a smaller amount within the first 10 Myr.}
\label{fig:radlnom}
\end{figure}

Fig.~\ref{fig:radlnom} depicts the evolution of Lagrangian radii in the unsegregated clusters. The evolution of Lagrangian radii is determined 
by three competing processes, two-body relaxation which decreases the core radii before core collapse and increases all radii after core collapse,
as well as gas expulsion and mass loss by stellar evolution which increase the cluster radii. The
relative importance of the different processes depends on the
initial half-mass radius and number of cluster stars. In the most compact clusters with $r_h=0.033$ pc, 
the relaxation time is short enough that they 
undergo core collapse before gas expulsion sets in. They later expand due to gas expulsion 
and mass loss from massive stars in the core and can expand by up to a factor 10 within 10 Myr. Interestingly, the
final radius reached after 10 Myr of evolution is the same for clusters starting with $r_h=0.033$ pc and $r_h=0.10$ pc. Hence the initial
radius cannot be determined from the one after 10 Myr for very compact clusters. In more extended clusters, the initial 
relaxation time is too long to allow significant mass segregation before gas expulsion sets in, so these clusters 
become mass segregated and go into core collapse only on a much longer timescale of several Myr. This is most clearly visible for the
$N=10^4$ stars, $r_h=1$ pc cluster which has the longest relaxation time of all clusters studied and for which the core is contracting  
after $T=5$ Myr. More compact clusters and clusters with lower particle numbers have developed some degree of mass segregation so that
the expansion due to mass-loss outweighs contraction due to two-body relaxation and the clusters always expand.

\begin{figure}
\begin{center}
\includegraphics[width=8.5cm]{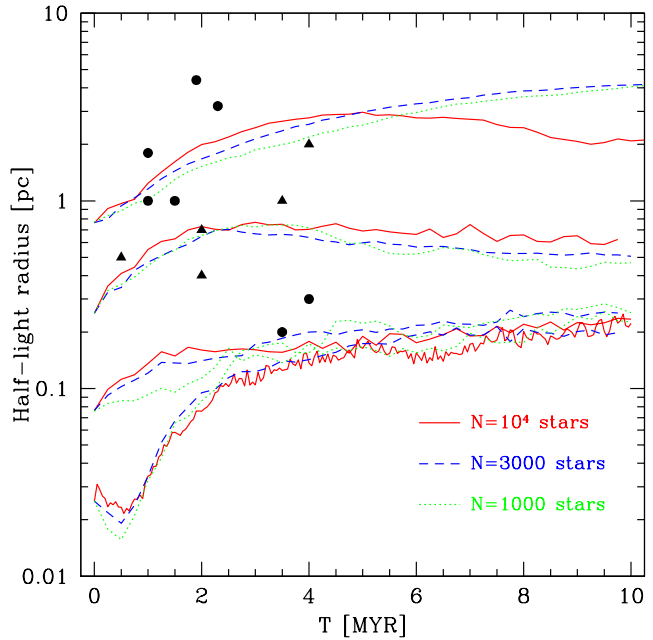}
\end{center}
\caption{Evolution of projected Lagrangian radii of bright stars for unsegregated clusters as a function of time. Lagrangian radii
increase due to gas expulsion, stellar evolution, and dynamical heating by binaries and decrease due to mass segregation of the
brightest stars into the cluster cores, with the role of the different processes depending on the starting condition. Filled triangles
show the positions of young open clusters from Portegies Zwart et al. (2010) while filled circles show the positions of additional
Galactic open clusters taken from the literature. The radii of observed clusters are most compatible with models starting with half-mass 
radii in the range 0.2 to 2 pc.}
\label{fig:radbnom}
\end{figure}

Fig.~\ref{fig:radbnom} depicts the evolution of projected half-light radius of bright stars in the unsegregated clusters. We have calculated
the Lagrangian radii from the positions of all stars with masses larger than 3 M$_\odot$ that have not yet turned
into compact remnants (mainly black holes given the short calculation times of our runs). The mass limit of
3 M$_\odot$ was chosen to allow a comparison with observations since in distant star clusters stars fainter than 
this often cannot be observed and contribute only a small fraction of the total cluster light. 
Filled triangles show the age and half-light radii of open clusters from Portegies Zwart et al. (2010), which are clusters in the
mass range $10^4$ M$_\odot < M_C < 10^5$ M$_\odot$. Filled circles   
show the location of additional open clusters with masses in the range $10^3$ M$_\odot < M_C < 10^4$ M$_\odot$ 
and which contain massive stars according to Weidner et al. (2010). The parameters of the clusters shown can be found in Table~2.  

It can be seen that Lagrangian radii of bright stars also mostly expand, except for clusters starting with half-mass radii
$r_h=0.33$ pc for which mass segregation causes a slight decrease at later times. The most compact clusters all reach 
radii of about $r_{hp}=0.2$ pc at T=10 Myr. This radius therefore represents a minimum radius for star clusters and there
are indeed no observed clusters smaller than it. Simulations starting with half-mass radii $r_h \le 0.1$ pc 
lead to radii significantly smaller than those of observed clusters. For the unsegregated clusters discussed here the
best match with observations is achieved for models starting with half-mass radii in the range $0.33$ pc $< r_h < 1$ pc.

\begin{figure}
\begin{center}
\includegraphics[width=8.5cm]{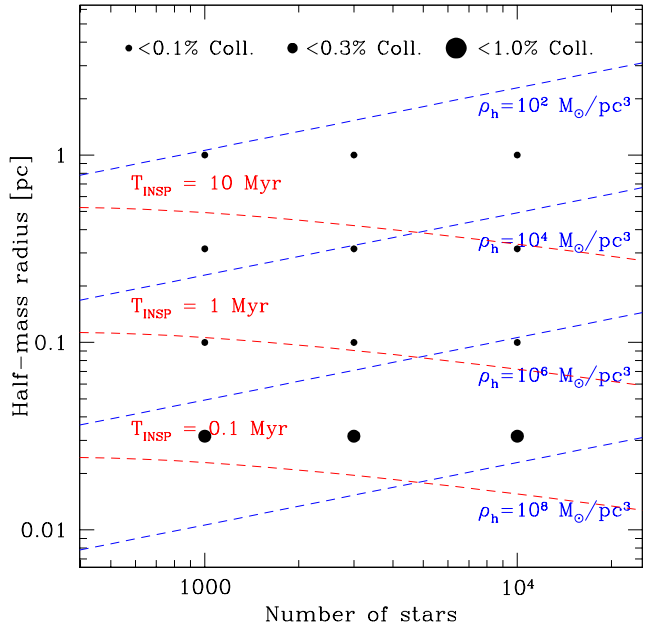}
\end{center}
\caption{Collision rate of stars for the unsegregated clusters. Collisions only play a role for initial densities larger than
$\rho_h=10^6$ M$_\odot$/pc$^3$ and inspiral times of a few 0.1 Myr or less. However, even in the most concentrated clusters, the number
of collisions are too small to populate a complete mass function up to 100 M$_\odot$, for which roughly the most massive 1\% of all stars 
have to collide with each other. Clusters with $r_h>0.$ pc, which lead to half-mass radii in agreement with
galactic open clusters have negligible number of collisions.}
\label{fig:coll}
\end{figure}

We next discuss the number of stellar collisions in the unsegregated clusters. According to \citet{spzmm02}, the inspiral time of a star 
of mass $m> \langle m \rangle$ is given by
\begin{eqnarray}
\nonumber T_{INSP} & = & 3.3 \frac{\langle m \rangle}{m} T_{rh} \\
  & = & 0.4 \frac{\sqrt{M_{C}} r_h^{3/2}}{\sqrt{G} m \ln{0.02 N}}
\end{eqnarray}
where $T_{rh}$ is the half-mass relaxation time of the cluster which we have taken from \citet{s87} and where $\langle m \rangle$ is the mean stellar mass. Significant number of collisions
are only to be expected after massive stars have spiraled into the cluster core, because low-mass stars have small radii and therefore
smaller cross sections for collisions.
Fig. \ref{fig:coll} depicts the number of stellar collisions in the unsegregated cluster models. In clusters less concentrated than
$r_h \approx$ 0.1~pc 
hardly any collisions occur. The reason is the low stellar density of these models in combination with the
large inspiral times of the massive stars. In clusters with half-mass radii larger than $r_h = 0.1$ pc, the inspiral times are 
larger than a few 0.1 Myr, meaning that massive stars do not have time to accumulate in the center while still in the accretion
phase. Instead, they reach the center only after a few Myr, at which point they have arrived already on the main-sequence and
have much smaller radii and collision cross sections. However, even in the most massive and concentrated cluster with $N=10^4$ stars and 
initial half-mass radius $r_h=0.033$ pc only 41 stars collide with each other. For a Kroupa mass function with stars up to 100 
M$_\odot$, 90 collisions between 10 M$_\odot$ stars are necessary to create the missing stars if the initial mass function only extends
up to 15 M$_\odot$. The number of collisions in the unsegregated models are therefore too small compared to the required number of collisions,
especially for clusters which have final half-light radii compatible with observed clusters, i.e. clusters starting with $r_h > 0.1$ pc.

We note that in the unsegregated runs most collisions happen after massive stars have reached their final mass, in the $N=10^4$ stars,
$r_h=0.033$ pc run for example only 3 collisions happen in the pre-MS phase while roughly half of the collisions happen after 1 Myr,
i.e. by the time the cluster has already become gas free and collisions would possibly be observable through their collision products or
flashes of bright emission. 

\begin{table}
\caption[]{Basic data of young star clusters containing high-mass stars with masses larger than $M>30$ M$_\odot$.}
\begin{center}
\begin{tabular}{lcccl}
\hline\hline
 Cluster Name &  Age & log Mass & Radius & Ref\\ 
  & [Myr] & [M$_\odot$] & [pc] & \\
Arches       &  2.0 &  4.3 & 0.4 & 1\\ 
NGC 3603     &  2.0 &  4.1 & 0.7 & 1\\
Quintuplet   &  4.0 &  4.0 & 2.0 & 1\\
Westerlund 1 &  3.5 &  4.5 & 1.0 & 1\\
Trumpler 14  &  0.4 &  3.6 & 0.5 & 1,2 \\
ONC          &  1.0 &  3.3 & 1.0 & 3,4 \\
NGC 637      &  4.0 &  3.2 & 0.3 & 3,5 \\ 
NGC 2244     &  1.9 &  3.8 & 4.4 & 3,6\\ 
NGC 6231     &  1.0 &  3.7 & 1.8 & 3,7 \\
NGC 6530     &  2.3 &  3.0 & 3.2 & 3,6\\
Westerlund 2 &  1.5 &  3.9 & 1.0 & 3,8 \\
DBS 2003     &  3.5 &  3.8 & 0.2 & 3,9 \\
\hline \hline
\end{tabular}
\end{center}

Notes: 1: Portegies Zwart et al. (2010), 2: Sana et al. (2010), 3: Weidner et al. (2010), 4: Hillenbrand \& Hartmann (1998), 
5: Yadav et al. (2008), 6: Chen et al. (2007), 7: Raboud (1999), 8: Ascenso et al. (2007), 9: Borissova et al. (2008)
\label{tabobscl}
\end{table}

\subsection{Mass-segregated clusters}
\label{sec:seg}

\begin{figure}
\begin{center}
\includegraphics[width=8.5cm]{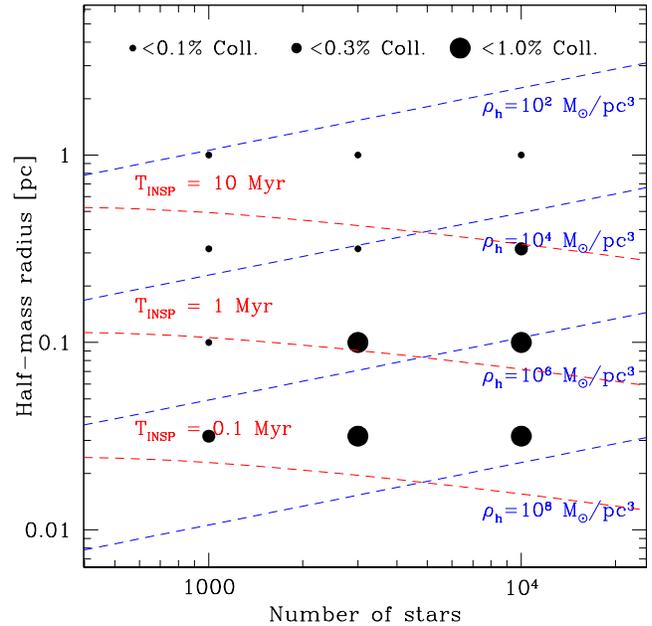}
\end{center}
\caption{Same as Fig. \ref{fig:coll} for the segregated clusters. Compared to the unsegregated case, the number of stellar collisions is
increased by about a factor 10. In clusters with initial densities larger than $\rho_h=10^6$ M$_\odot$/pc$^3$, the number of collisions
is high enough to allow the formation of complete main-sequence of high-mass stars.}
\label{fig:collmseg}
\end{figure}

We next discuss the evolution of the mass-segregated clusters. In the segregated clusters energy and final
mass of stars were correlated such that proto-stellar cores which will become the highest mass stars have 
the lowest energies. The high mass stars therefore
form in the cluster center and do not have to spiral into the center by dynamical friction. Furthermore, the cluster core contracts 
during the accretion process due to mass increase, further facilitating stellar collisions. Table~1 and Fig. \ref{fig:collmseg} show
the number of collisions in the segregated models. It can be seen that the number of collisions which occur in the runs are now about 
a factor 10 higher than in the unsegregated models. In particular in the most concentrated models starting with $r_h=0.033$ pc,
the number of collisions is now sufficiently high to build up a main-sequence of high-mass stars. A closer look at the data also
shows that in the mass segregated models the collisions happen at earlier times on average, for the star cluster with $N=10^4$ and $r_h=0.033$ pc
cluster, for example, roughly half of all collisions happen in the pre-MS stage.
We note that these results are not in contradiction to results obtained by \citet{ardietal2008}, who found that primordial mass segregation 
does not lead to a significant increase in the number of stellar collisions, since in our simulations the increase in the number of 
collisions comes to a large part from an 
increase in the central density as a result of the mass accretion of the heavy stars. In contrast, Ardi et al. started their simulations with all stars 
being on the main sequence and the same mass density for segregated and unsegregated clusters, meaning that the stellar number density in the
mass segregated clusters was much lower than in the unsegregated models.

\begin{figure}
\begin{center}
\includegraphics[width=8.5cm]{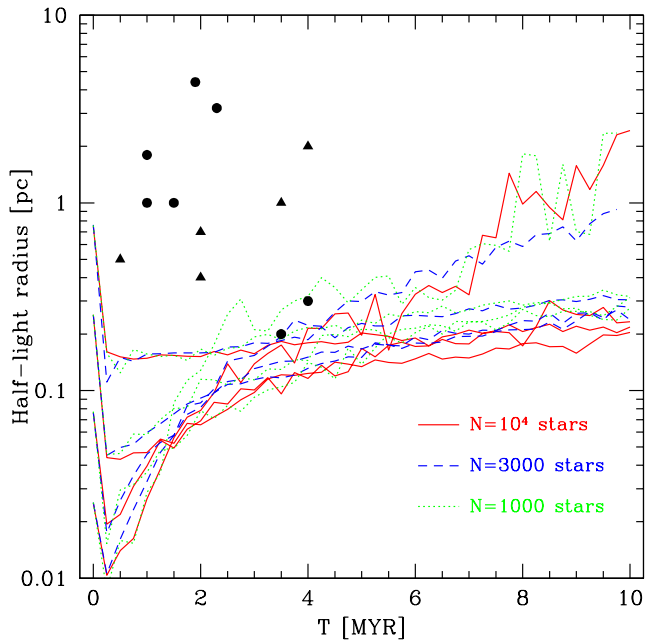}
\end{center}
\caption{Evolution of projected Lagrangian radii of bright stars for the mass-segregated clusters as a function of time. Lagrangian radii
of bright stars decrease during the first few $10^5$ yr due to mass accretion of massive stars, which decreases the size of the cluster
core. Later radii increase due to dynamical evolution and stellar evolution mass loss. The projected half-light radii stay 
significantly below the observed radii of young star clusters. If star clusters form mass segregated, the starting radii must 
be larger than several pc in order to fit observed cluster radii after a few Myr.}
\label{fig:radbmseg}
\end{figure}

Fig.\ \ref{fig:radbmseg} depicts the evolution of the projected radius containing
50\% of bright stars as a function of time. Due to the initial mass segregation, all models have significantly smaller half-light light
radii than the non-segregated models of the previous section. The final half-light radii are also significantly smaller than in
the previous section, despite some expansion 
due to gas expulsion, stellar evolution, and dynamical
heating by binary stars. The projected half-light radii of bright stars are for all models also well below the measured radii of young 
open clusters. It therefore is unlikely that any of the mass segregated models represents the starting condition of any observed 
Milky Way cluster. Starting with even larger cluster radii would be a way to get a better fit to observed cluster radii, however this
would again reduce the number of collisions to too small levels. Strong cluster expansion on a timescale of a few $10^5$ yrs after the most 
massive stars have formed would be needed to bring the cluster radii into agreement with observed cluster radii. This may be achieved
through a very small star formation efficiency of the order of 10\% or less, which however is much smaller than observed star
formation efficiencies (Lada \& Lada 2003). It therefore seems that primordial mass segregation and a shrinking cluster core
are not a way to get a sufficiently high number of collisions for realistic clusters. 

\begin{figure}
\begin{center}
\includegraphics[width=8.5cm]{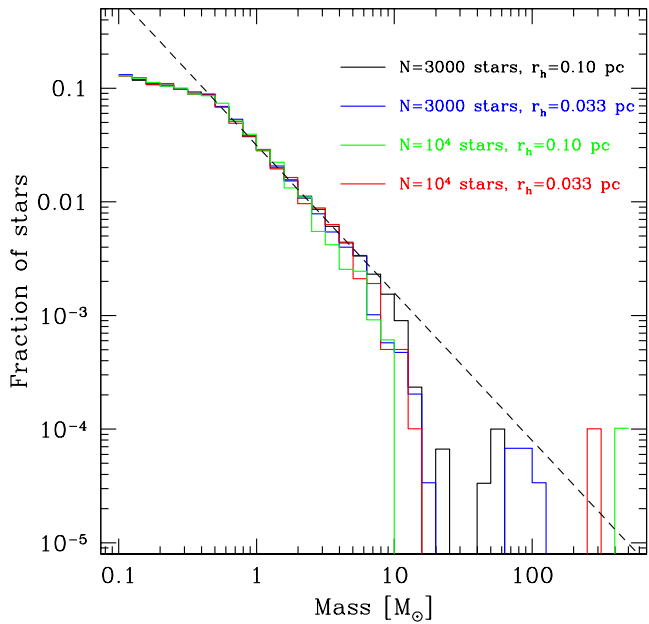}
\end{center}
\caption{Final mass function of stars at $T=10$ Myr in the 4 mass segregated models with the highest number of collisions. Dashed line shows
the expected number of stars for a Kroupa (2001) IMF extending up to 100 M$_\odot$. Stellar collisions do not lead to the build-up of
such a mass function, instead they mainly lead to the formation of single, very massive stars. The behavior is similar to the
runaway merging scenario found for main-sequence stars in dense star clusters.} 
\label{fig:mfuncmseg}
\end{figure}

An additional problem 
for the hypothesis that high-mass stars form by stellar collisions is illustrated in Fig. \ref{fig:mfuncmseg}. It shows
the final mass function of stars in the four mass segregated models which have the largest number of collisions. All 
models show a sharp drop in the number of stars at $m=15$ M$_\odot$, meaning that collisions were not able to build a 
sufficient number of stars with $m>15$ M$_\odot$. Instead, the main effect
of collisions was the formation of a single massive star in each cluster, inspection of the runs shows that typically more than
90\% of all collisions involve the same star. This behavior is similar to what \citet{spzmm02} found for collisions between main-sequence
stars in dense star clusters: The most massive star has a large radius and also attracts other stars via gravitational focusing. As
a result it has by far the highest cross section for collisions of all stars in the core and once it has grown beyond a certain
critical mass, all further collisions happen nearly exclusively with this star.
Instead of building many massive stars in the mass range 15 M$_\odot < m <$ 100 M$_\odot$, there is runaway growth of a single star 
only. This star can reach masses with more
than 100 M$_\odot$ in some of our runs. Although these runs are probably too concentrated to lead to realistic star clusters,
runaway collisions might be possible in more massive star clusters with larger radii. Collisions will also be more common if stars 
more massive than the mass limit of 15 M$_\odot$ chosen here can form directly through gas accretion since higher mass stars
have larger cross sections. Runaway collisions between massive
stars during the pre-MS phase could therefore lead to the formation of very massive stars in the densest star clusters, 
like the massive WN5-6 stars with masses up to 320 M$_\odot$ recently found in several massive star clusters by \citet{cetal10}. 
They might also contribute to the formation of intermediate-mass black holes in dense star clusters \citep{spzetal04}.
These results are similar to results recently obtained by \citet{mc10}, who also found that collisions between massive
proto-stars lead to the formation of one or two extremely massive objects instead of a smooth filling of the top end of the
mass spectrum. 

\subsection{Influence of the mass accretion rate and binary stars}
\label{sec:maccbin}

\begin{figure}
\begin{center}
\includegraphics[width=8.5cm]{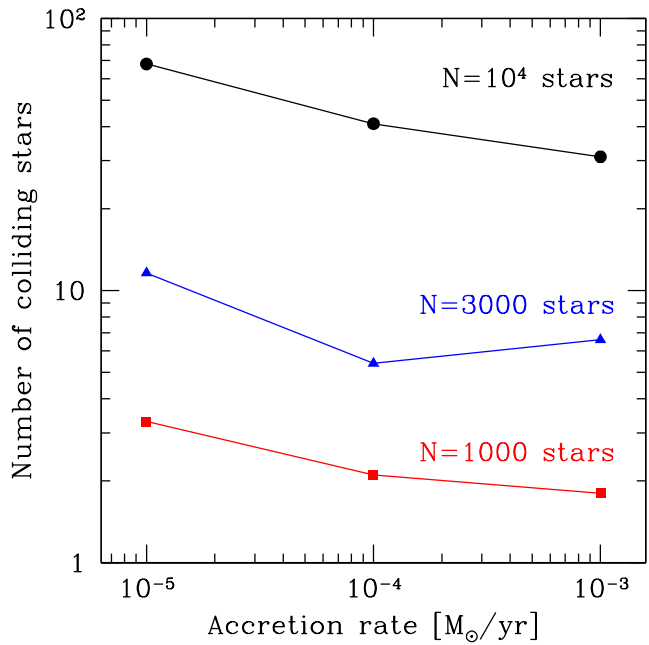}
\end{center}
\caption{Number of stellar collisions as a function of the mass accretion rate in runs starting with half-mass radius $r_h=0.033$ pc. The number
of collisions changes by less than a factor of three between the different runs since e.g. for high accretion rates, the larger radii of stars in 
the pre-main sequence phase are compensated for by the smaller time spend in the pre-main sequence phase.} 
\label{fig:macc}
\end{figure}

Fig.~\ref{fig:macc} depicts the influence of the mass accretion rate on the number of collisions occurring in a cluster
within the first 10 Myr. All clusters start from a half-mass radius of $r_h = 0.033$ pc and are unsegregated. According to
eq.~2, stars accreting with a larger rate are more extended, which will lead to a higher number of collisions. However,
larger accretion rates also lead to shorter pre-main sequence times which decreases the number of collisions.
Fig.~\ref{fig:macc} shows that both effects nearly cancel each other
The accretion
rate changes only by a factor of two despite the fact that the accretion rate varies between $10^{-5}$ M$_\odot$/yr to $10^{-3}$ M$_\odot$/yr.
This is too small to significantly change the results of the previous paragraphs.

Another possible effect which was not considered so far is the influence of binary stars. It is well known that most massive stars
reside in binary or multiple systems (Garc{\'{\i}}a \& Mermilliod 2001, Lada 2006, Sana et al. 2007). Stellar binaries have a 
larger cross section for collisions than single stars, which could
enhance the number of collisions. We therefore performed a number of simulations containing primordial binaries. We assumed an initial binary
fraction of 50\% in these runs and that all high-mass stars reside in binaries. Binary masses were chosen such that stars were ordered 
according to their mass and the components of each binary were drawn from consecutive stars in this list, i.e. the first binary contained 
the two most massive stars. Hence, the stellar binary fraction is 100\% for the high-mass stars, while all low mass stars are single.
Although being a strong simplification, our adopted binary distribution reflects the drop in binary fraction towards late spectral types
seen for stars in the galactic disc (Lada 2006).

The semi-major
axis of the binaries were chosen randomly in $\log r$ between a minimum radius three times as large as the sum of the radii reached
at the end of the accretion phase and a maximum radius which was set equal to 100 AU. For simplicity, we assumed in our runs that 
the semi-major axis of each binary remains constant during the accretion phase and increased the stellar velocities of the components 
to avoid a shrinkage of the binary.

\begin{figure}
\begin{center}
\includegraphics[width=8.5cm]{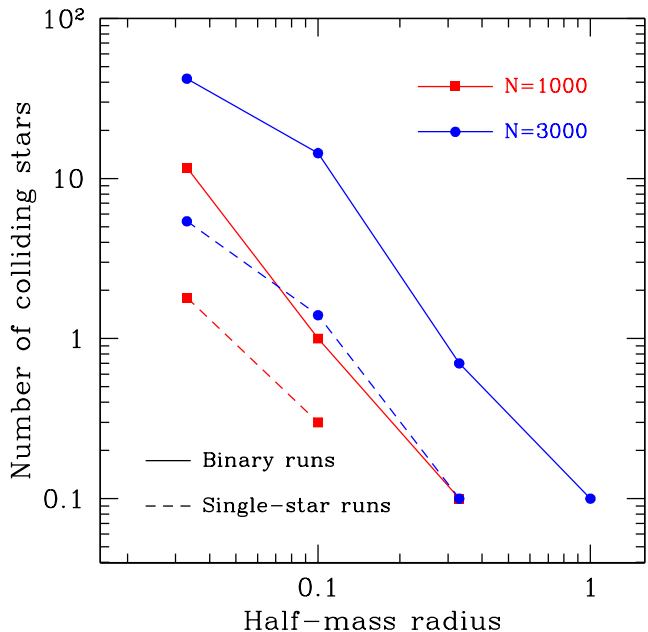}
\end{center}
\caption{Number of stellar collisions in runs containing 50\% binaries. Binaries increase the number of collisions by a factor of 5 to 10
compared to the corresponding single star runs. However, for clusters with initial half-mass radii $r_h>0.1$ pc, this is still not
large enough to create a sufficient number of high-mass stars.} 
\label{fig:bin}
\end{figure}

Fig.~\ref{fig:bin} shows the resulting number of collisions in runs with binary stars. The number of collisions is large in very
compact clusters with $r_h = 0.33$ pc.  The projected half-light radii of these clusters after 10 Myrs are similar to 
those of clusters without binaries and still smaller than about 0.3 pc (see Table~1). Such clusters therefore still end up too compact compared 
to observed open clusters. In more extended clusters the number of collisions is still not high enough to allow the build-up
of a complete main sequence of massive stars. It is therefore likely that even in the presence of binaries, stellar collisions
do not play a significant role for the formation of massive stars. A definite answer to this question can, however, only be made
if a wider range of binary distributions is explored.   

\section{Conclusions}
\label{sec:concl}

We have performed $N$-body simulations of the pre-main sequence evolution of stars in 
stellar clusters, taking account gas accretion, primordial gas expulsion, and collisions between stars. Our simulations show 
that it is very unlikely that all high-mass stars with masses $m>20$ M$_\odot$ form from the collisions of lower mass stars. The
reason for this is twofold: First, the necessary number of collisions between massive stars only occurs for central densities
around $10^8$ M$_\odot$/pc$^3$, implying initial half-mass radii $r_h<0.1$ pc for clusters of a few thousand stars. Such clusters
remain highly concentrated within the first 10 Myr of their evolution, despite expansion due to gas expulsion and stellar evolution mass
loss, and therefore lead to clusters which are significantly more concentrated than known open clusters with O and B type stars which
generally have radii around 1 pc. Our simulations show that the observed radii of young open clusters imply initial radii in the
range 0.2 to 1 pc for most of them.      

Second, even if a sufficient number of collisions occurs, this will normally lead to the formation of single runaway stars with extremely 
large masses instead of the build-up of the observed high-mass IMF, a result that was recently also obtained by \citet{mc10} through
direct $N$-body simulations. The reason is the large cross section for collisions and gravitational focusing of massive
stars. The number of collisions cannot be increased sufficiently by considering different mass accretion rates or binary stars. 
It therefore appears unlikely that massive stars form by
collisions. Instead, mass growth by accretion through an accretion disc seems the more likely
explanation for the formation of massive stars.

The question arises if other initial conditions not studied in this paper, like dynamically cool models or clumpy 
stellar distributions could change the number of collisions sufficiently. Dynamically cool models are  
unlikely to lead to dramatically higher collision rates. Although the central density may increase considerably during 
the initial contraction phase, this time period is very short as the cluster will virialise within a few crossing times 
and then evolve along the same tracks as the models considered here. Since our simulations typically last for 10s to 1000s 
of crossing times, an initial collapse phase lasting a few crossing times would not significantly change the outcome of our 
simulations. We suspect the same to be true for clumpy initial conditions. Again any potential contraction phase would be 
short compared to the overall evolutionary timespan considered, and in addition clumps that are very small or very compact 
will quickly dissolve by three body encounters. However, a more definite answer will require a detailed parameter study of 
the dynamical evolution of initially non-relaxed dense clusters. This is beyond the scope of this first assessment of the problem.

In clusters more massive than the ones studied here, or in clusters containing more massive stars, collisions between pre-main sequence 
stars might become important. It is attractive to speculate that they might lead to the formation of a few extremely massive stars, 
like the Pistol star in the Quintuplet star cluster \citep{fetal98, netal09}, or the massive WN5-6 stars recently found in several 
massive clusters by \citet{cetal10}, and therefore ultimately to the formation of intermediate-mass black holes with masses
up to a few hundred M$_\odot$.

\section*{Acknowledgements}

We thank Paul Crowther and an anonymous referee for valuable comments on the manuscript. 
H.B. acknowledges support from the German Science foundation through a Heisenberg Fellowship and from the
Australian Research Council through Future Fellowship grant FT0991052.
R.S.K.\ acknowledges financial support from the {\em Landesstiftung Baden-W{\"u}rttemberg}
via their program International Collaboration II (grant P-LS-SPII/18) and from the German
{\em Bundesministerium f\"{u}r Bildung und Forschung} via the ASTRONET project STAR FORMAT (grant 05A09VHA).
R.S.K. furthermore gives  thanks for subsidies from the {\em Deutsche Forschungsgemeinschaft}  under
grants no.\ KL 1358/1, KL 1358/4, KL 1359/5, KL 1358/10, and KL 1358/11, as well as from a Frontier
grant of Heidelberg University sponsored by the German Excellence Initiative. R.S.K. also thanks the
KIPAC at Stanford University and the Department of Astronomy and Astrophysics at the University of
California at Santa Cruz for their warm hospitality during a sabbatical stay in spring 2010. KIPAC is supported in part by the U.S. Department of Energy contract no. DE-AC-02-76SF00515.

\label{lastpage}

\end{document}